\documentclass[12pt]{iopart}
\usepackage{graphicx}

\begin{document}

\title{Consequences of nonconformist behaviors in a continuous opinion model}
 
\author{Allan R. Vieira$^{1}$, Celia Anteneodo$^{2,3}$, Nuno Crokidakis$^{1}$}

\address{
$^{1}$ Instituto de F\'{\i}sica, Universidade Federal Fluminense, Niter\'oi/RJ, Brazil \\
$^{2}$ Departamento de F\'isica, PUC-Rio, Rio de Janeiro/RJ, Brazil \\
$^{3}$ National Institute of Science and Technology for Complex Systems, Brazil}

\ead{allanrv@if.uff.br, celia.fis@puc-rio.br, nuno@if.uff.br}

\begin{abstract}
\noindent
We investigate opinion formation in a kinetic exchange opinion model, 
where opinions are represented by numbers in the real interval  $[-1,1]$ and  
agents are typified by the  individual degree of conviction about the opinion that they support. 
Opinions evolve through pairwise interactions governed by competitive positive and negative couplings, 
that promote imitation and dissent, respectively. 
The model contemplates also another type of nonconformity such that agents can occasionally 
choose their opinions independently of the interactions with other agents. 
The steady states of the model as a function of the  parameters that describe  conviction, 
dissent and  independence are analyzed, with particular emphasis on the emergence of extreme opinions. 
Then, we characterize the possible ordered and disordered phases and the occurrence or suppression 
of phase transitions that arise spontaneously due to the disorder introduced by the heterogeneity 
of the agents and/or their interactions.

\end{abstract}

Keywords: Dynamics of social systems, Collective phenomena, Computer simulations, Critical phenomena

\maketitle

\section{Introduction}

\qquad
Statistical physics furnishes diverse tools for the study 
of human social dynamics. 
Despite the apparent dissimilarities between social and  physical systems, 
many social processes present a phenomenology that resembles that 
found, for instance, in the physics of frustrated or disordered 
materials~\cite{loreto_rmp,galam_book,sen_book,galam_review,pawel}. 
This is because, despite the high heterogeneity of the individuals, 
and their interactions,  not all details are relevant for the 
emergence of collective patterns. Collective behaviors make  social systems interesting for 
the physicist and, reciprocally, the physicist might contribute with a new perspective 
to the comprehension of social phenomena. 

One of the basic ingredients to be taken into account 
for modeling people's interactions is imitation, or social contagion. 
In fact, imitation is  observed in diverse social contexts, 
such as in the dynamics of language learning or decision making. 
The recurrent conformity to the attitudes,  opinions or decisions  
of other individuals, or groups of individuals, has 
led to the formulation of models based on social contagion as 
the primary rule of opinion dynamics, e.g., the voter~\cite{voter}, Sznajd~\cite{sznajd} and 
majority rule~\cite{majority} models, to give just a few examples. 
However, aside from imitation, individuals also 
dissent and resist to be influenced, in several ways.  

In the present work, we study the effect of nonconformity attitudes   
through a kinetic exchange model~\cite{sen_book,sen,lccc,biswas}. 
Within this modeling, the influences that an individual exerts over 
another are modulated by a coupling strength. 
The coupling strength between connected individuals typically takes positive values but also, 
with certain probability,  it can adopt negative ones.  
Negative values represent negative influences that, instead of imitation, promote dissent. 
Notice that this kind of dissent is not a characteristic of the individual but of the tie 
(or link) between each pair of individuals. 

There are also other types of nonconformity, which are not associated to the links 
   but to the individuals (or nodes in the network of contacts), as taken into account in 
several models~\cite{galam,lalama,sznajd_indep1,sznajd_indep2,sznajd_indep3,nuno_indep,javarone,bagnoli}. 
One of these types  is  anticonformity~\cite{willis,nail}. 
An anticonformist actively dissents from other people's opinions, 
which is the case contemplated, for example, by  Galam's contrarians~\cite{galam_book,galam_review}. 
Actually,  although these anticonformists defy other people's opinions or the group norm, 
they are similar to conformists in the sense that they take into account other's opinions too. 
A different kind of nonconformity is independence, such that the individual tends to resist the influences of other 
agents or groups of agents, ignoring their choices in the adoption process. 
It can be thought either as an attribute of the agent, that acts as independent with certain probability, 
or an attitude that any individual can assume with certain frequency. It is in the latter sense 
that we will incorporate independence into the model. 

Another important ingredient for modeling  opinion dynamics is 
conviction or persuasion (a kind of stubbornness or resistance 
to change mind)~\cite{galam_book,sen_book,sen,deffuant2,nuno_celia,brugna,nuno_pmco_jstat,xiong,nuno_jstat,
sibona,marlon}. 
We will consider the heterogeneity of the individuals in that respect. 
Each agent will be characterized by a parameter that measures its level of conviction 
about the opinion it supports. This parameter, typically defined positive, 
will be allowed to take also negative values  to represent  volatile individuals 
that change mind easily.

Heterogeneities and disorder can act on  opinion dynamics as  stochastic drivings 
able to promote a phase transition,   
playing  the role of a source of randomness or noise similar 
to a social temperature~\cite{lalama,sznajd_indep2,sznajd_indep3,nuno_indep,nuno_celia,victor}. 
We will focus on the impact of all the abovementioned sources of disorder on the steady state 
distribution of opinions in the population, 
and investigate the occurrence of nonequilibrium phase transitions.

% ###########################################################################

\section{Modeling}

\qquad 
Our model belongs to the class of kinetic exchange opinions models~\cite{sen,lccc,biswas}. 
We consider a fully-connected population of size $N$ participating in a public debate. 
Each agent $i$ in this artificial society has an opinion $o_{i}$.  
Most models of opinion dynamics deal with a  discrete state 
space~\cite{loreto_rmp,galam_review,pawel}, 
which may be enough to tackle certain problems that involve binary or several particular choices. 
However,  to investigate, for instance, the emergence of extreme opinions, 
it seems to be  more suitable to represent opinions by means of a continuous 
variable,  to reflect the  possible shades of peoples' attitudes about a given 
subject~\cite{deffuant2,sibona,marlon,fan,hk,deffuant,deffuant3,lorenz,coda,wu}.  
Then, we will consider a continuous state space, where opinions can take values  
in the real range $[-1,1]$. 
Positive (negative) values indicate that the position is favorable (unfavorable) to the topic under discussion. 
Opinions tending to $1$ and $-1$ indicate extremist individuals. 
Finally, opinions near $0$ mean neutral or undecided agents. 

At a given step  $s$, the following microscopic rules control the opinion dynamics:

\begin{enumerate}

\item We choose two random agents $i$ and $j$. 

\item With probability $q$, the agent $i$ acts as independent. 
In this case, the agent chooses a position about the subject under discussion, 
i.e., $o_{i}(s+1)$ is randomly selected from a uniform distribution $[-1,1]$, 
independently of the current states $o_{i}(s)$ and $o_{j}(s)$. 
That is, as commented above, we consider that independence is not an attribute of the individual 
but an attitude that any individual  
can occasionally assume with certain probability.  In that opportunity,  individuals choose their own position (state) 
independently of the other individuals~\cite{sznajd_indep1,sznajd_indep2,sznajd_indep3,nuno_indep}. 

\item Otherwise, with probability $1-q$, the agent $i$ acts as a partially conformist individual, 
and will be influenced by agent $j$ by means of a kinetic exchange. 
In this case, the opinion of agent $i$ will be updated according to  the rule \cite{lccc,biswas}
\begin{equation} \label{eq1}
o_{i}(s+1) = c_i o_{i}(s)+\mu_{ij}\,o_{j}(s) ~,
\end{equation}
where  $c_i$ and $\mu_{ij}$ are real parameters that measure  the level of conviction of agent $i$, and  
  the strength of the influence that agent $i$ suffers from agent $j$, respectively. 
The opinions are restricted to the range $[-1,1]$. 
Therefore, whenever Eq.~(\ref{eq1})  yields $o_{i}>1$ ($o_{i}<-1$),  it will, actually,  
lead the opinion to the extreme $o_{i}= 1$ ($o_{i}=-1$), as considered in Refs.~\cite{sen,lccc,biswas}. 
This re-injection of the opinions into the interval $[-1,1]$ introduces 
a nonlinearity in the mapping given by Eq.~(\ref{eq1}), that becomes linear by parts.
Smoothing this nonlinearity, for instance through  $o_i(s+1) \to \tanh[o_i(s+1)]$, 
is not expected to affect the results significantly, as 
will be discussed below in connection with the outcomes of simulations.

Typically, the strength $\mu_{ij}$ is positive, meaning agreement,  
but it can also take negative values to represent ties that induce dissent. 
The fraction of negative influences is given by a parameter $p$, 
such that $\mu_{ij}$ is uniformly distributed either in $[-1,0]$ or in  $[0,1]$,  
with probabilities $p$ and $1-p$, respectively.

Parameter $c_i$ is also typically positive, representing the weight or contribution that 
the current opinion of agent $i$ has, with respect to that of other agents,  
in the opinion formation process. 
The larger $c_i$, the less other individuals contribute to mold 
the opinion of agent $i$. 
Sufficiently large values of $c_i$ (larger than 1 for the current choice of the set $\mu_{ij}$)   
mimic intransigence or stubbornness,  such that the opinion of agent $i$ will not evolve.
Parameter $c_i$ can also take negative values,  to represent  volatile individuals that are not 
persistent in their positions, that is, 
have a propensity to change mind.   
This spontaneous change must be distinguished from the change of opinion in independent behavior described above. 
Due to the minus sign, an agent  with negative conviction will always change side, 
adopting a new opinion which is the opposite of the previous one. 
Differently, in the independent case, since the new opinion is chosen at random,  
the individual opinion will not necessarily change sign but can take any value in the spectrum. 

The fraction of negative convictions will be controlled by a parameter $w$, such 
that  $c_{i}$ is uniformly distributed  either in $[-1,0]$ or in $[0,1]$, with 
probabilities $w$ and $1-w$, respectively.

\end{enumerate}

The unit of  time  is defined by the application of the above-mentioned steps $N$ times. 
Concerning the random nature of the convictions $c_i$ and the couplings $\mu_{ij}$, 
which are assumed to be uncorrelated, we  consider quenched (frozen) variables, 
as far as opinion formation supposedly occurs in a time scale much faster than the changes in agents' attributes and couplings. 
 
 We will consider populations of  $N$  fully-connected individuals, situation which 
corresponds to a mean-field limit. The initial state of the system is assumed to be fully disordered,  
that is,  at the beginning of the dynamics, each individual has an opinion drawn from the uniform  distribution  
in the range $[-1,1]$.

In the following, we will describe separately two distinct cases characterized by: 
(i)  the existence of competitive positive/negative interactions among pairs of agents, while 
convictions are homogeneous (with $c_i=1$, $\forall i$), 
and (ii) the heterogeneity of agents' convictions, which interactions are always positive 
($\mu_{ij}>0$, for all $i,j$). 
In both cases, the noise introduced by independent attitudes is included.

\section{Methods}

\qquad
We calculate the parameter $O$ given by
\begin{equation} \label{eq2}
O = \left\langle \frac{1}{N}\left|\sum_{i=1}^{N} o_{i}\right|\right\rangle ~, 
\end{equation}
where $\langle\, ...\, \rangle$ denotes average over  disorder or configurations, 
computed at the steady states. 
Notice that $O$ is a kind of order parameter that 
plays the role of  the ``magnetization per spin'' in  magnetic systems. 
It is sensitive to the unbalance between positive and negative opinions. 
A state with a large value of $O$ ($O\simeq 1$) means that an extremist position reached consensus. 
Intermediate values indicate:  (i)  the dominance of either one of the extreme opinions, 
(ii) that opinions are moderate but one of the sides wins, or (iii) a combination of both. 
All these states can be identified with ordered ones in the sense that the debate has a clear 
result (favorable or unfavorable), be extremist or moderate.  
A small value ($O\simeq 0$) indicates a symmetric distribution of opinions:  
(i) polarization such that opposite opinions balance, 
(ii) the dominance of very moderate or undecided opinions around the neutral state, 
or (iii) a combination of both. 
In all symmetric cases,   the debate will not have a clear winner position. 
In that sense the  collective state can be identified with a disordered one.

We  consider also the fluctuations $V$  (or ``susceptibility'') of  parameter $O$, 
\begin{equation} \label{eq3}
V =  N\,(\langle O^{2}\rangle - \langle O \rangle^{2})  \,, 
\end{equation}
%onesss
and the Binder cumulant $U$~\cite{binder}, defined as
\begin{equation} \label{eq4}
U   =   1 - \frac{\langle O^{4}\rangle}{3\,\langle O^{2}\rangle^{2}} \,.
\end{equation}

All these quantities will be used to characterize the phase transitions between ordered and 
disordered phases. Additionally, those phases will be described by means of 
the pattern that the distribution of opinions presents.
% ##########################

\section{Results: Competitive interactions and independence}
\label{case1}

\begin{figure}[h]
\begin{center}
\vspace{0.5cm}
\includegraphics[width=0.3\textwidth,angle=270]{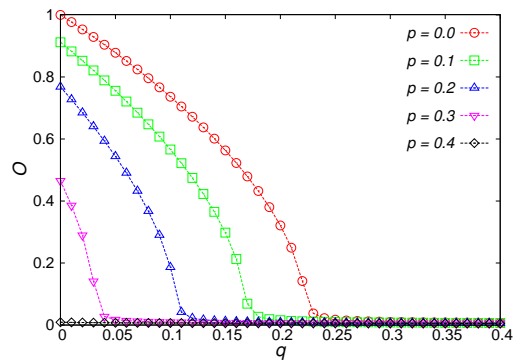}
\end{center}
\caption{Order parameter $O$ versus the independence probability $q$, 
for several values of the fraction $p$ of negative interaction strengths, 
with $c_i=1$ for all $i$. 
One can observe transitions at different critical points $q_{c}(p)$, 
but the maximal value of $O$ decreases with $p$ and the 
transition is eliminated for  sufficiently large values of $p$. The population size is $N=10^{4}$ 
and data are averaged over $100$ simulations.}
\label{fig1}
\end{figure}

\qquad
Let us focus first on the effect of competitive interactions (with a fraction $p$ of negative couplings) and 
independence (which occurs with probability $q>0$), by studying  
the case of  homogeneous agents with $c_i=1$ for all $i$.
In Figure~\ref{fig1} we exhibit $O$ as a function of the independence parameter $q$, 
for different values of  $p$. 
As observed in the figure,  independence makes the system  undergo  a phase transition, 
an  effect which is typical of  social temperatures~\cite{lalama,sznajd_indep1}.

%%%%%%%%%%%%%%%%%%%%%%%%%%%%%%%%%%%%%%%%%%%%%%%%%%%%%%%%%%%%%%%%%%%%%%%%%%
%%%%%%%%%%%%%%%%%%%%%%%%%%%%%%%%%%%%%%%%%%%%%%%%%%%%%%%%%%%%%%%%%%%%%%%%%%%

The transition occurs even when couplings are all positive ($p=0$),   
then the mere presence of independence  leads the system to undergo  a phase transition. 
A similar effect was observed in the discrete version of the present model, 
considering only opinions $o=-1$, $0$, $1$, 
although the critical point $q_c(p=0)$ has a  different value (1/4 in the discrete case~\cite{nuno_indep}).

When we introduce negative interaction strengths ($p>0$),  states characterized by $O=1$, 
meaning consensus of extreme opinions, become unlikely (see Figure~\ref{fig1}). 
Moreover,  the critical values $q_{c}$ decrease with $p$ and, above sufficiently large $p$, 
the system becomes  disordered  for any $q$. 
The critical value $p_{c}(q=0)\approx 0.34$ is in accord  with that found in Ref.~\cite{biswas}.

\begin{figure}[h]
\begin{center}
\vspace{0.5cm}
\includegraphics[width=0.3\textwidth,angle=270]{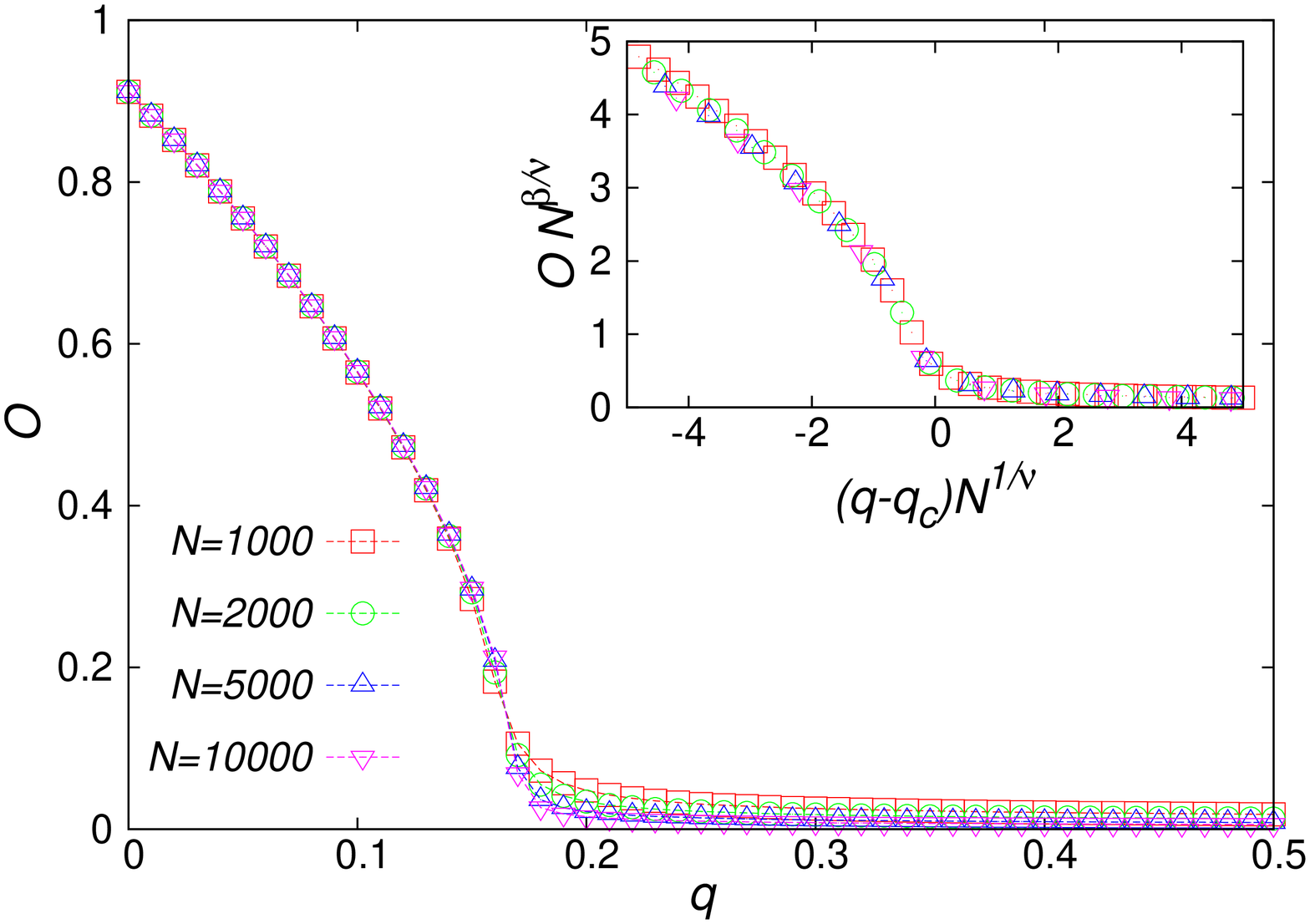}
\hspace{0.5cm}
\includegraphics[width=0.3\textwidth,angle=270]{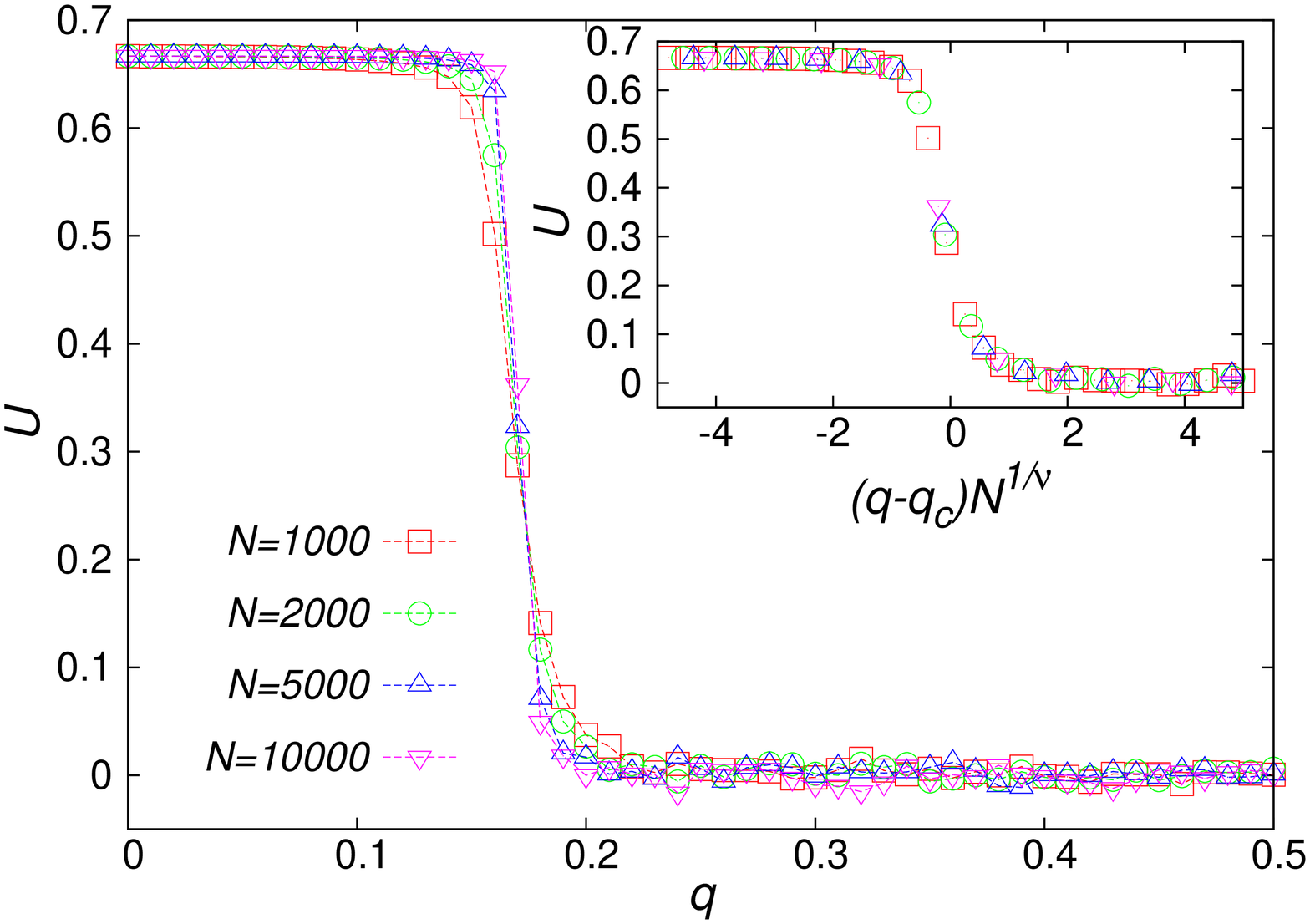}
\\
\vspace{1.0cm}
\includegraphics[width=0.3\textwidth,angle=270]{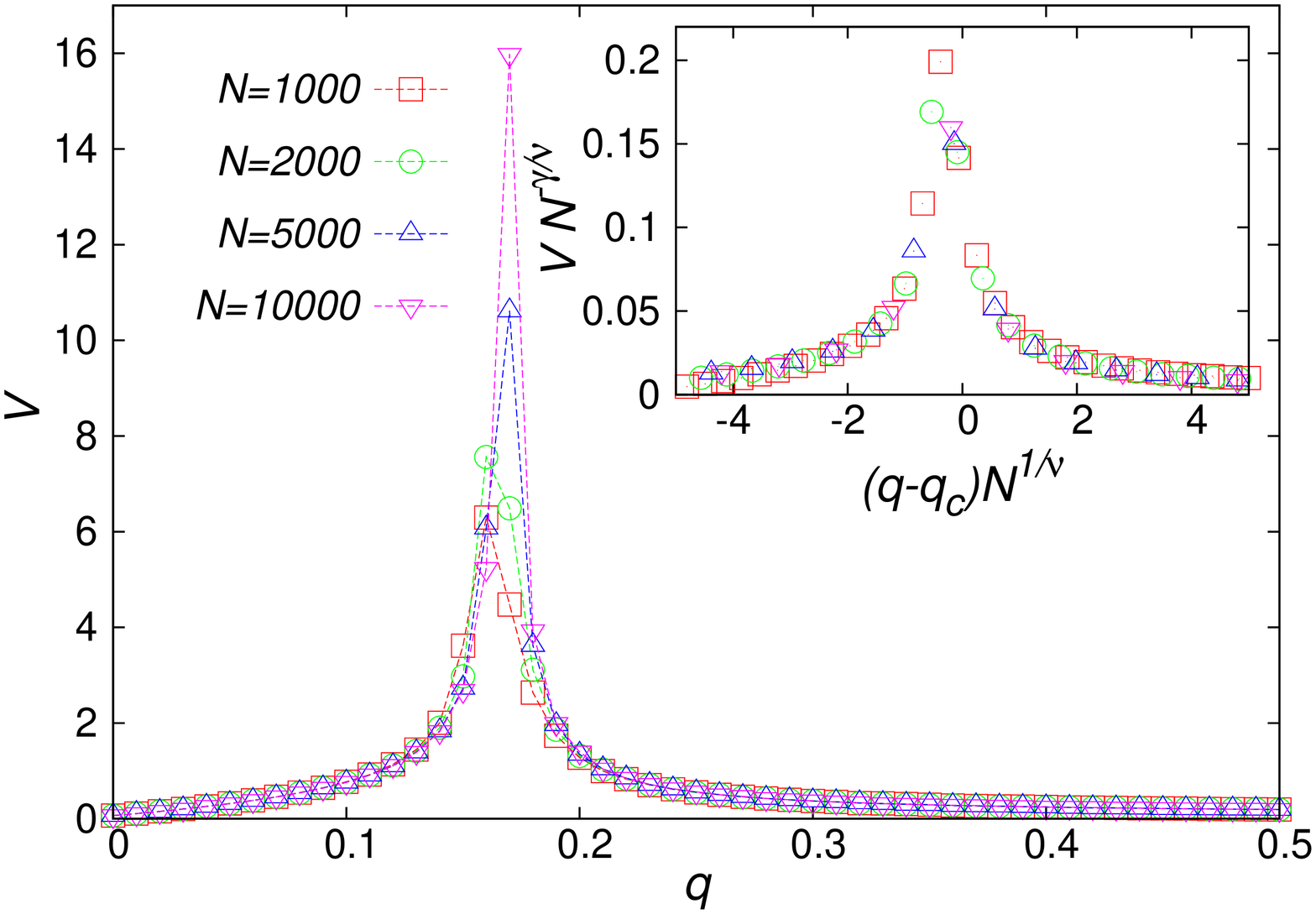}
\end{center}
\caption{Finite-size scaling analysis of the transition as a function of $q$, for $p=0.1$ and 
$c_i=1$ for all $i$. 
The best data collapse  was found for $q_{c}\approx 0.173$, $\beta\approx 1/2$, 
$\gamma\approx 1$ and $\nu\approx 2$.}
\label{fig2}
\end{figure}

To estimate the critical points $q_{c}(p)$, we performed a finite-size scaling (FSS) analysis. 
As a typical example, we exhibit in Figure~\ref{fig2} the results of FSS for $p=0.1$. 
The critical exponents  are $\beta\approx 1/2$, $\gamma\approx 1$ and $\nu\approx 2$. 
Thus, as expected, the universality class known for the mean-field implementation of the model when $q=0$~\cite{biswas} 
is not altered by the presence of the microscopic disorder introduced by independence.

%%%%%%%%%%%%%%%%%%%%%%%%%%%%%%%%%%%%%%%%%%%%%%%%%%%%%%%%%%%%%%%%%%%%%%%%%%
%%%%%%%%%%%%%%%%%%%%%%%%%%%%%%%%%%%%%%%%%%%%%%%%%%%%%%%%%%%%%%%%%%%%%%%%%%%

\begin{figure}[h]
\begin{center}
\vspace{0.5cm}
\includegraphics[width=0.3\textwidth,angle=270]{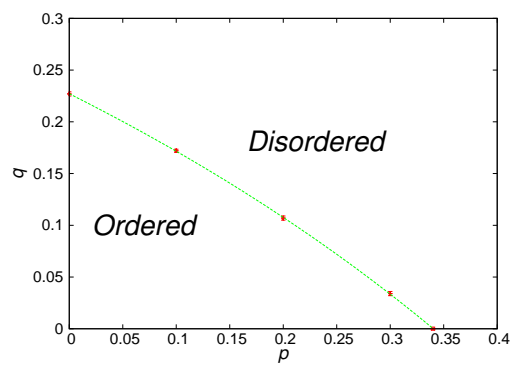}
\end{center}
\caption{Phase diagram in the plane $q$ (probability of independence) versus $p$ 
(probability of negative interactions), when $c_i=1\,, \forall i$. 
The symbols are the numerical estimates of the transition points, whereas the dashed line 
is given by Eq.~(\ref{eq5}). The error bars were estimated from the FSS analysis.
}
\label{fig3}
\end{figure}

The FSS analysis also provides the critical values $q_{c}(p)$, 
that allow to build a phase diagram in the plane $q$ versus $p$, depicted in  Figure~\ref{fig3}. 
The boundary separates the ordered and disordered phases. 
In the ordered phase, for $q<q_{c}(p)$, 
one of the sides (positive or negative opinions) 
will win the debate, while in the disordered phase, there will be balance of opposite opinions and/or 
dominance of moderate ones. 

Following the analytical expression found for the discrete kinetic exchange opinion 
models~\cite{nuno_celia},  namely a quotient of two first order polynomials in $p$, 
we propose  a qualitative description of the phase boundary through
\begin{equation}\label{eq5}
q_{c}(p)=\frac{a\,p+b}{c\,p+1}  ~,
\end{equation}
\noindent
which for fitting parameters $(a,b,c) \simeq (-0.667,0.227,-0.671)$ gives 
a heuristic description of the phase boundary, 
in good agreement with the critical points obtained from FSS analysis, as shown in Figure~\ref{fig3}. 
The extreme critical values $q_{c}(p=0)\approx 0.227$ and $p_{c}(q=0)\approx 0.341$   
are given respectively by  $b$ and $-b/a$. 
The latter value is in agreement with  the estimate found in Ref.~\cite{biswas}, which deals with such limiting case.

%%%%%%%%%%%%%%%%%%%%%%%%%%%%%%%%%%%%%%%%%%%%%%%%%%%%%%%%%%%%%%%%%%%%%%%%%%
%%%%%%%%%%%%%%%%%%%%%%%%%%%%%%%%%%%%%%%%%%%%%%%%%%%%%%%%%%%%%%%%%%%%%%%%%%%

%%%%%%%%%%%%%%%%%%%%%%%%%%%%%%%%%%%%%%%%%%%%%%%%%%%%%%%%%%%%%%%%%%%%%%%%%%
\begin{figure}[t]
\begin{center}
\vspace{0.5cm}
\includegraphics[width=0.3\textwidth,angle=270]{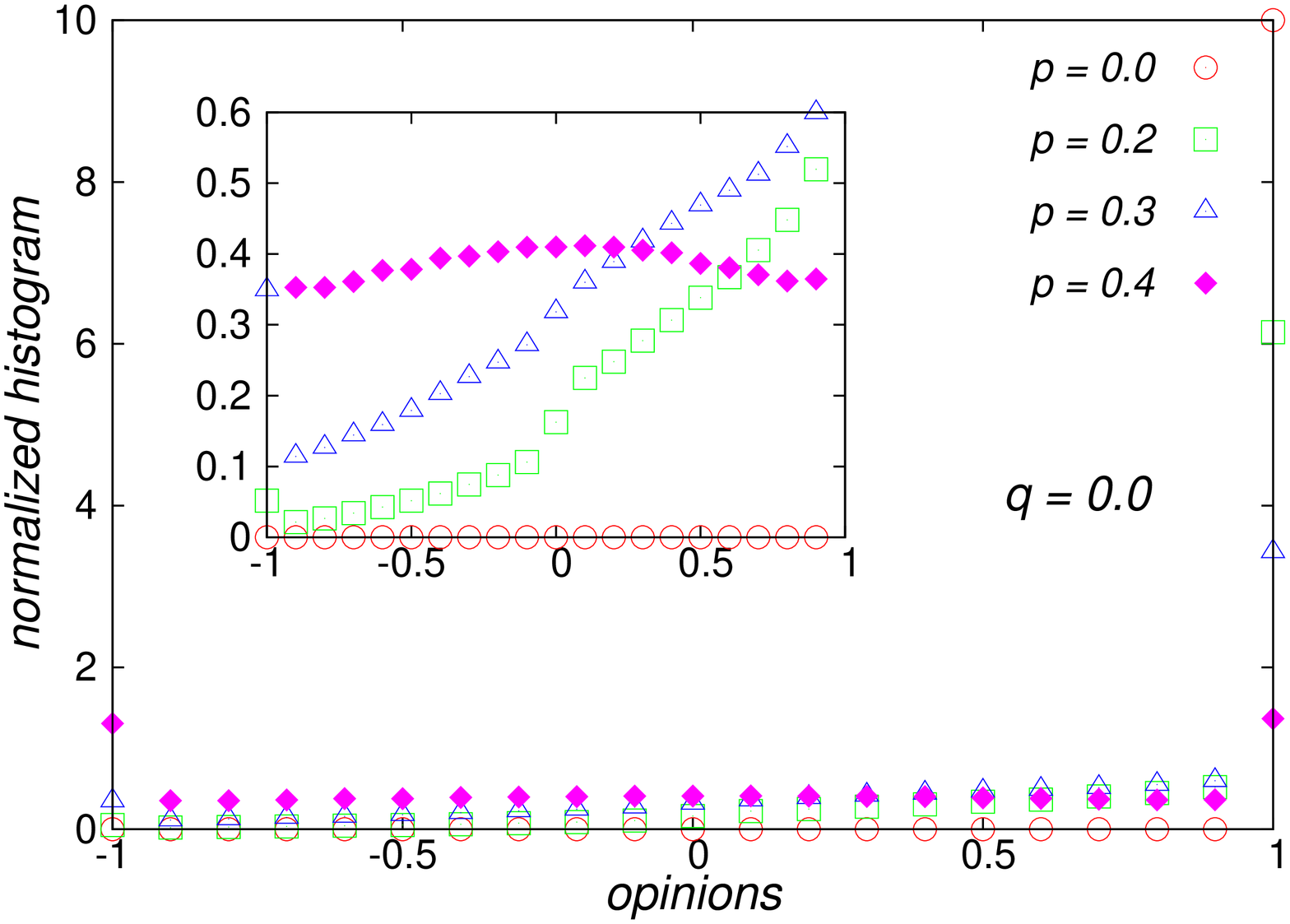}
\hspace{0.5cm}
\includegraphics[width=0.3\textwidth,angle=270]{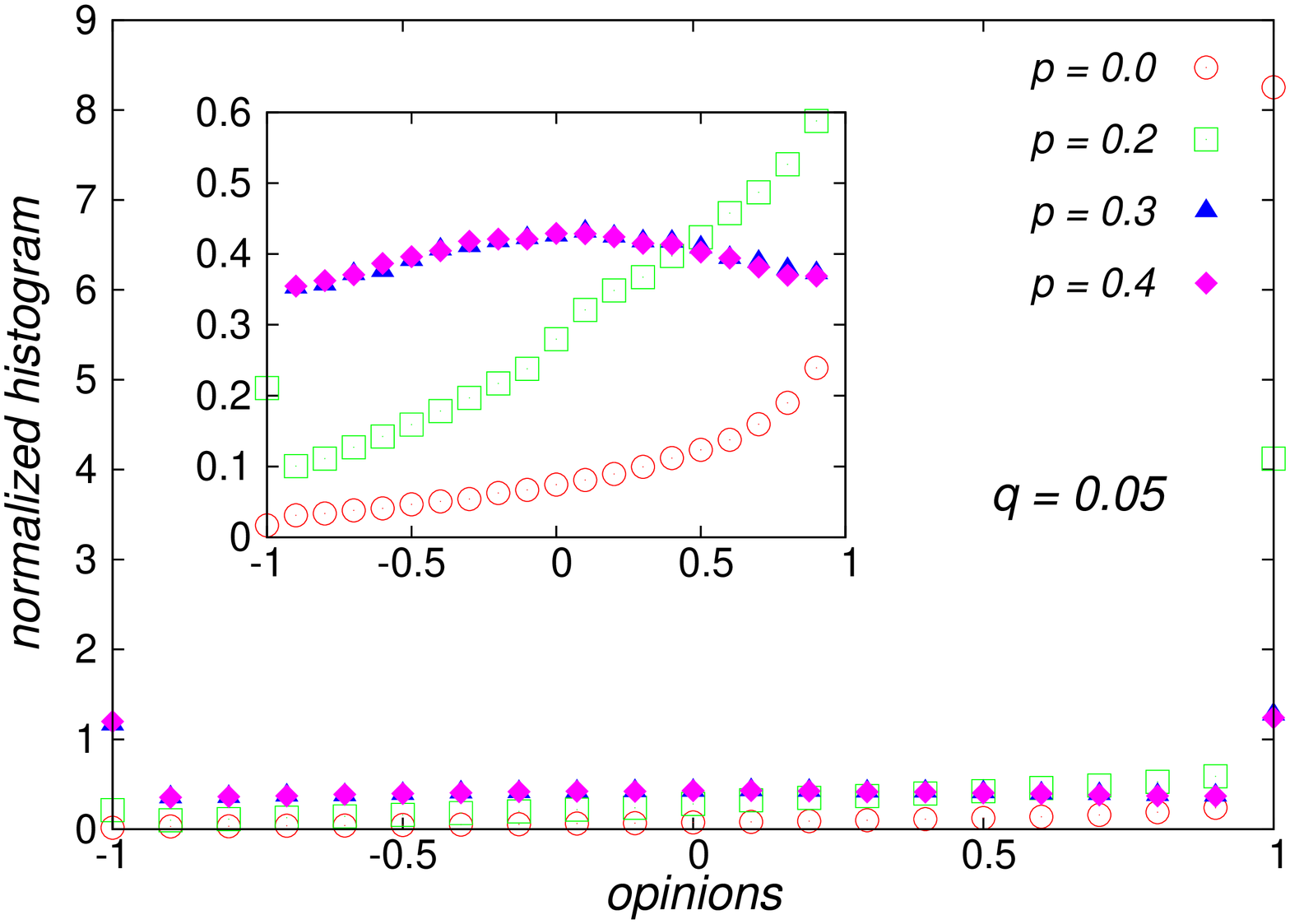}
\\
\vspace{1.0cm}
\includegraphics[width=0.3\textwidth,angle=270]{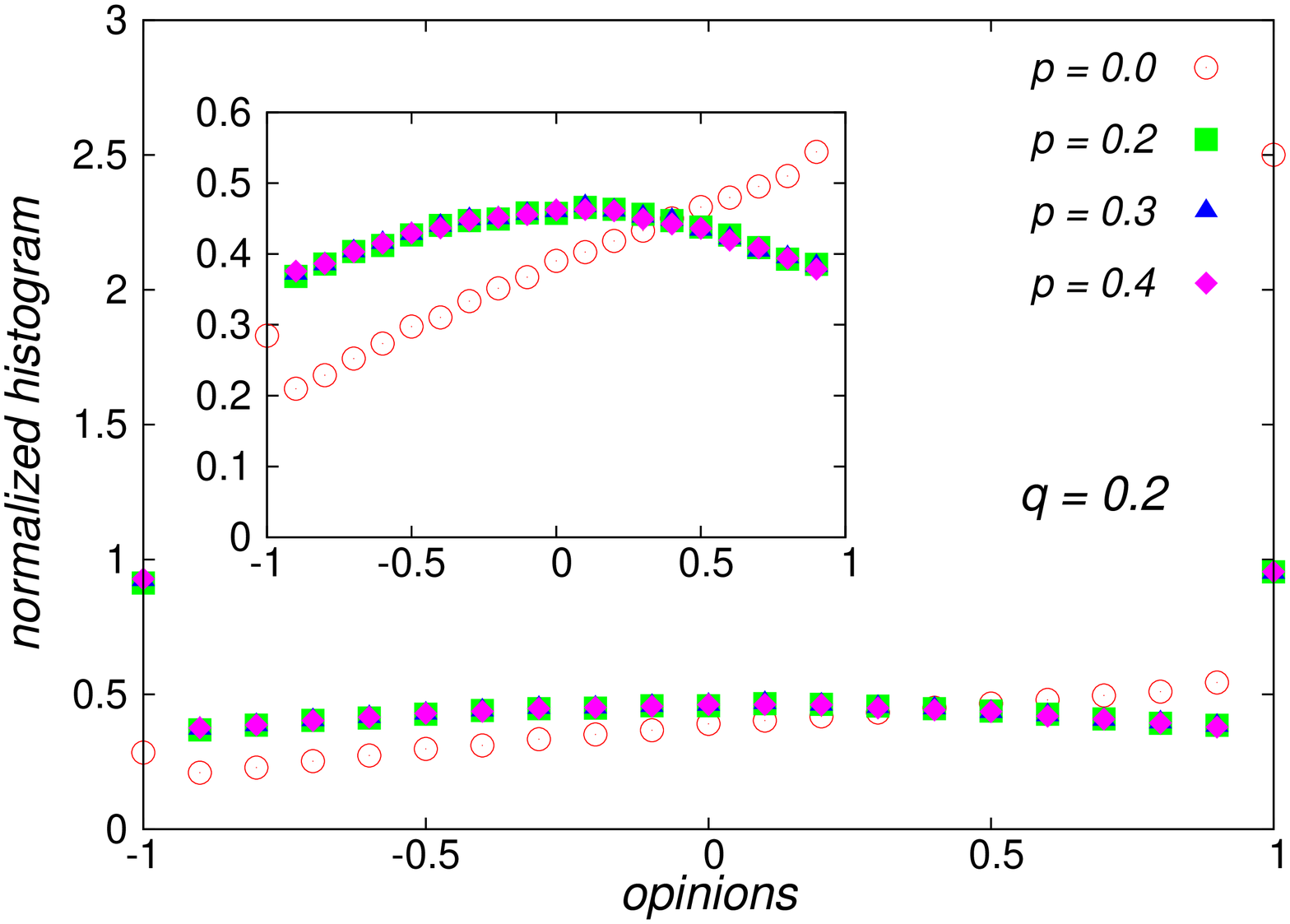}
\end{center}
\caption{Normalized histograms of opinions,  at the steady state,  for $q=0.0$ (top, left pannel), $q=0.05$  (top, right pannel) and $q=0.2$ (bottom), for several values of $p$, when $c_i=1\,, \forall i$. 
Empty (full) symbols represent ordered (disordered) states. In the insets we show a zoom of the main frames, excluding the extremist agents with majority opinions. The population size is $N=10^{4}$, and data are accumulated over $100$ independent simulations, as explained in the text.
}
\label{fig4}
\end{figure}
%%%%%%%%%%%%%%%%%%%%%%%%%%%%%%%%%%%%%%%%%%%%%%%%%%%%%%%%%%%%%%%%%%%%%%%%%%%

Since there is  ambiguity in the interpretation of the values of parameter $O$, 
we complemented the previous analysis,  computing the distribution of opinions in the population at the steady states of the model. 
In Figure~\ref{fig4}, we exhibit the outcomes for $q=0$, $q=0.05$ and $q=0.2$, at several values of $p$. 
Each normalized histogram is obtained from $100$ independent simulations, 
for population size $N=10^{4}$. 
When there is unbalance of positive and negative opinions,   
we arbitrarily selected  simulations with dominance of positive opinions to build the histograms. 
In this way, each histogram is representative of each single realization but with improved statistics. 
Instead, if we would have chosen the simulations with predominantly negative outcomes, the distribution would be 
the symmetric counterpart of those shown in Figure~\ref{fig4}.  
%a typical single realization of the dynamics, for population size $N=10^{4}$. 
First, we notice that there is  condensation of opinions at the extreme values $|o|=1$. 
This is a consequence of the type of nonlinearity introduced in Eq.~(\ref{eq1}). 
If a smoother form were used instead, then there will be a spread of the extreme values in the condensates, 
which will not affect significantly the number of individuals that can be identified as those who are extremists. 

For each frame (fixed value of $q$), we observe the following scenario.
Both the winner and loser sides  present  a bulk of moderate individuals as well as a condensation  
 at $o=\pm 1$ (extremists).
When $p$ is small enough so that the system becomes ordered, 
 unbalance between positive and negative opinions emerges,   
consistently with a non null order parameter. 
When $p$ becomes too large,  it is not possible the formation of  an ordered phase, where one of the sides dominates, 
then the distribution becomes flat, with the coexistence of all opinions 
(except for the symmetrical condensation in the extreme states $|o|=1$),  
as can be seen in the insets of Figure~\ref{fig4}.

Moreover, we see that the number of extremists (opinions $|o|=1$) decreases with increasing values of $p$,  
as well as with increasing $q$.   
That is, the inclusion of dissent interactions, as well as the growth of independent attitudes, 
inhibit extremism growth, leading to a more homogeneous distribution of opinions, 
which is associated to the disordered phase in 
the diagram of Figure~\ref{fig4}.

% ##########################

\section{Results: Heterogeneous convictions and independence}
\label{case2}

\qquad
In this section we consider  independence together with the heterogeneity of  the convictions $c_i$, 
which can take negative values. 
Recall that we assume that $c_{i}$ are (quenched) random variables that  
are uniformly distributed in $[-1,0]$ with probability $w$ and  uniformly distributed 
in $[0,1]$ with the complementary probability $1-w$. 
Hence, the parameter $w$ denotes the fraction of negative convictions. 
Positive convictions mean that the agent is sure about the adopted belief 
and therefore contributes in Eq.~(\ref{eq1}) to maintain the current opinion. 
A conviction close to zero reflects uncertainty about the belief and high susceptibility to other 
people's opinions  (conformist behavior), while a negative conviction contributes to form an indifferent or 
an opposite  position to the current one. 

When all the couplings are $\mu_{ij}=1$, the curves for the order parameter $O$ vs $q$,   
are qualitatively similar to those shown in  Figure~\ref{fig1} (after the substitution  $p \to w$), although the 
critical values are not the same. 
Moreover the distribution of opinions are also similar. 
Hence the cases (i) $c_i=1$ for all $i$, and variable $p$, and (ii) $\mu_{ij}=1$ for all $i,j$ with variable $w$, are in 
some way equivalent. 
Since  case (ii) does not present a  new phenomenology with respect to case (i), 
then we will exhibit, instead of $\mu_{ij}=1$,  
the results for a more realistic case, where all the interactions are positive but random, 
uniformly distributed in the range $[0,1]$, which corresponds to  $p=0$.

\begin{figure}[h]
\begin{center}
\vspace{0.5cm}
\includegraphics[width=0.3\textwidth,angle=270]{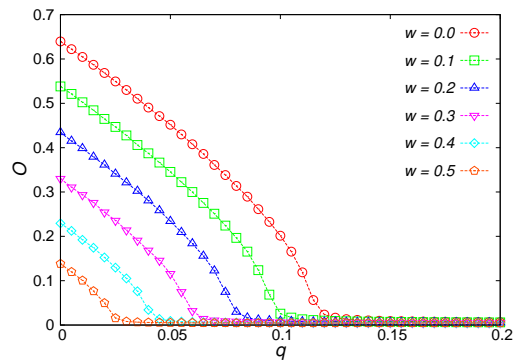}
\end{center}
\caption{Order parameter $O$ versus the independence probability $q$, 
for several values of the fraction $w$ of negative convictions, when $p=0$.
One can observe order-disorder transitions at different points $q_{c}(w)$, 
but the transition is suppressed for  sufficiently large values of $w$. 
The population size is $N=10^{4}$ and data are averaged over $100$ simulations. 
 }
\label{fig5}
\end{figure}

In Figure~\ref{fig5} we exhibit the behavior of the order parameter 
as a function of $q$ for several values of $w$. 
Although the curves are different from those in Figure~\ref{fig1}, 
the effect of convictions is rather similar to that introduced by  
negative interactions, in the sense that the increase of $w$, as well as the increase of $p$, 
both tend  to disorder the system. However, the population does not reach the state $O=1$  anymore, 
differently to the case $p=0$ of Figure~\ref{fig1}. 

%%%%%%%%%%%%%%%%%%%%%%%%%%%%%%%%%%%%%%%%%%%%%%%%%%%%%%%%%%%%%%%%%%%%%%%%%%
%%%%%%%%%%%%%%%%%%%%%%%%%%%%%%%%%%%%%%%%%%%%%%%%%%%%%%%%%%%%%%%%%%%%%%%%%%%

\begin{figure}[h]
\begin{center}
\vspace{0.5cm}
\includegraphics[width=0.3\textwidth,angle=270]{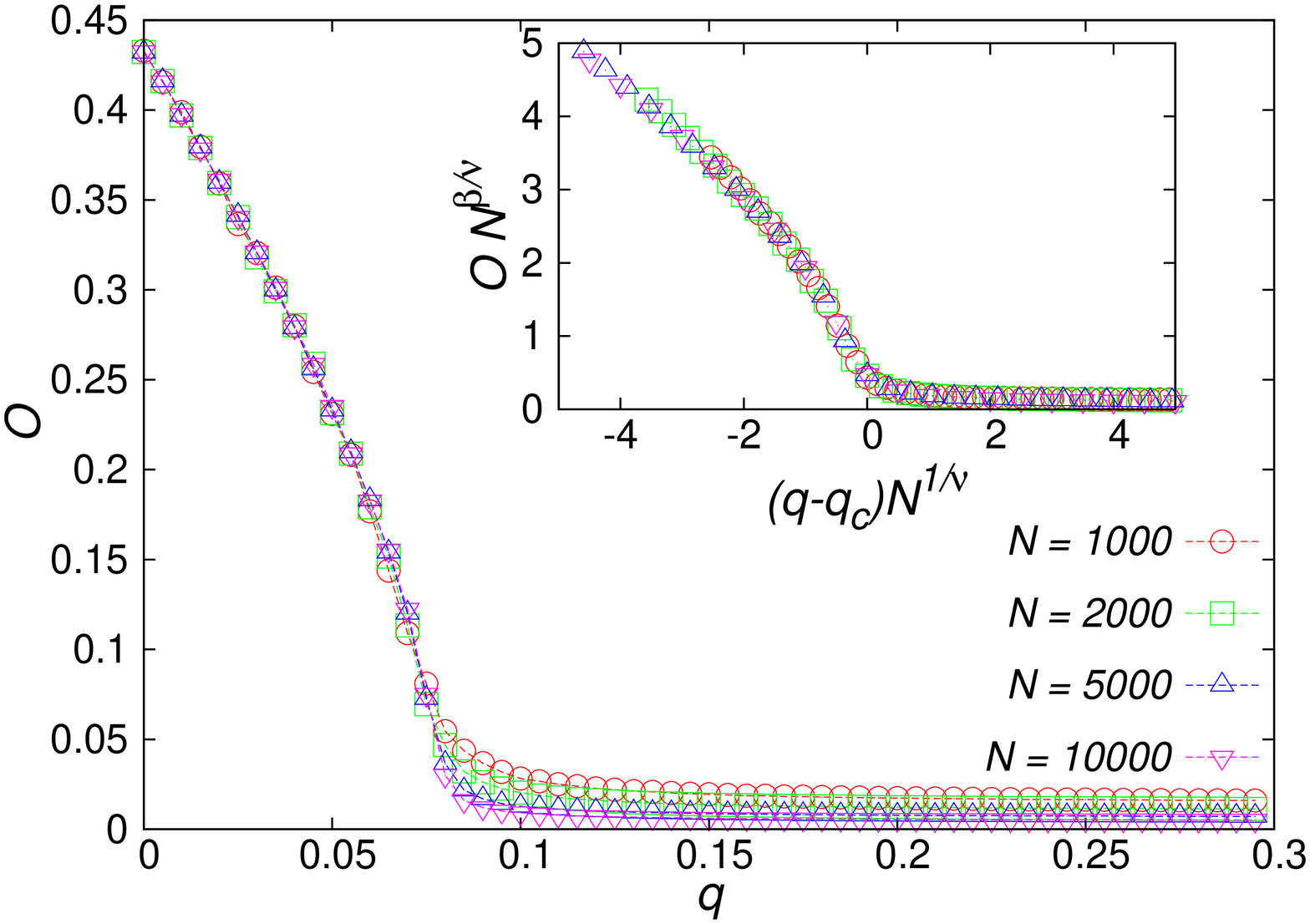}
\hspace{0.5cm}
\includegraphics[width=0.3\textwidth,angle=270]{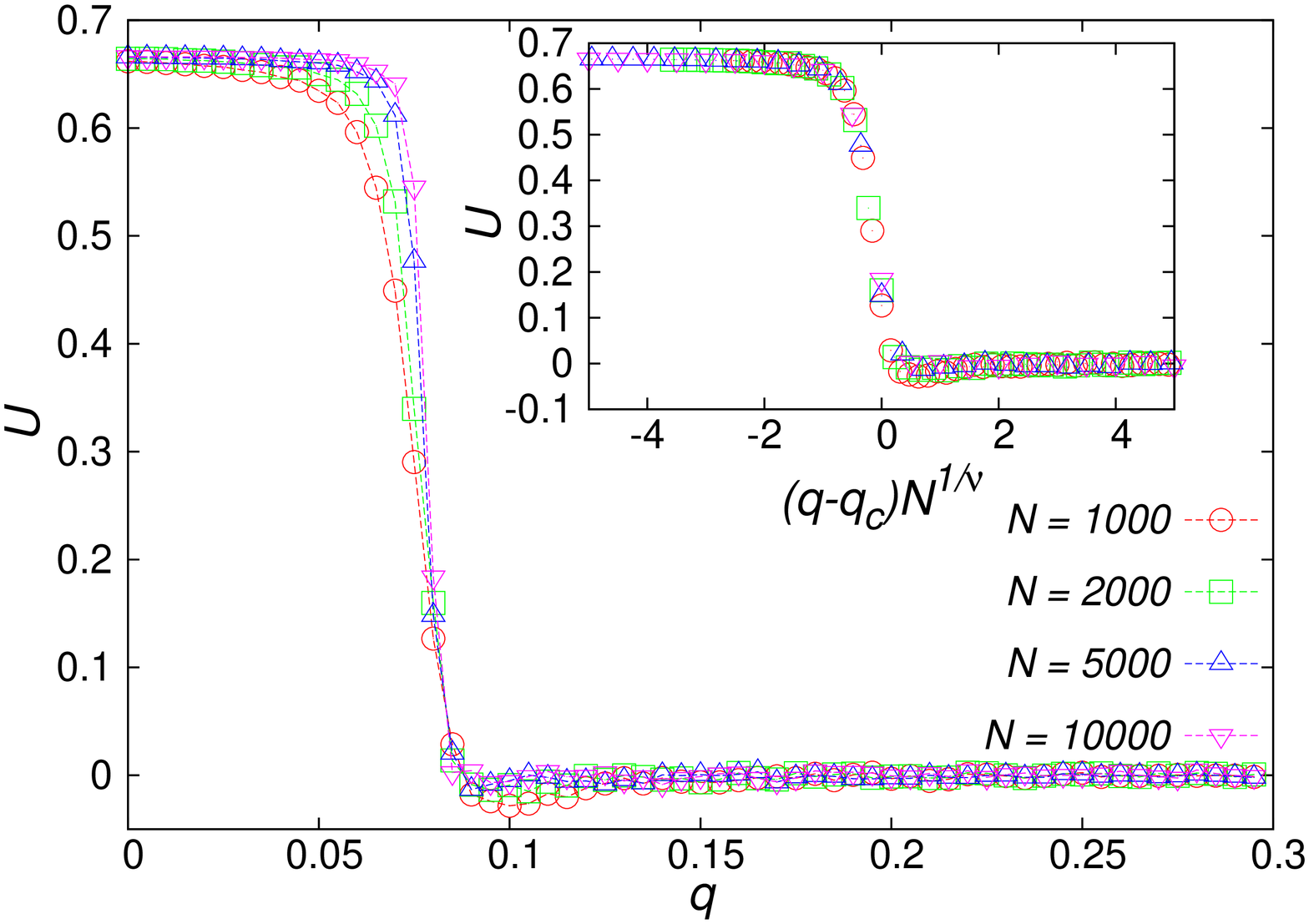}
\\
\vspace{1.0cm}
\includegraphics[width=0.3\textwidth,angle=270]{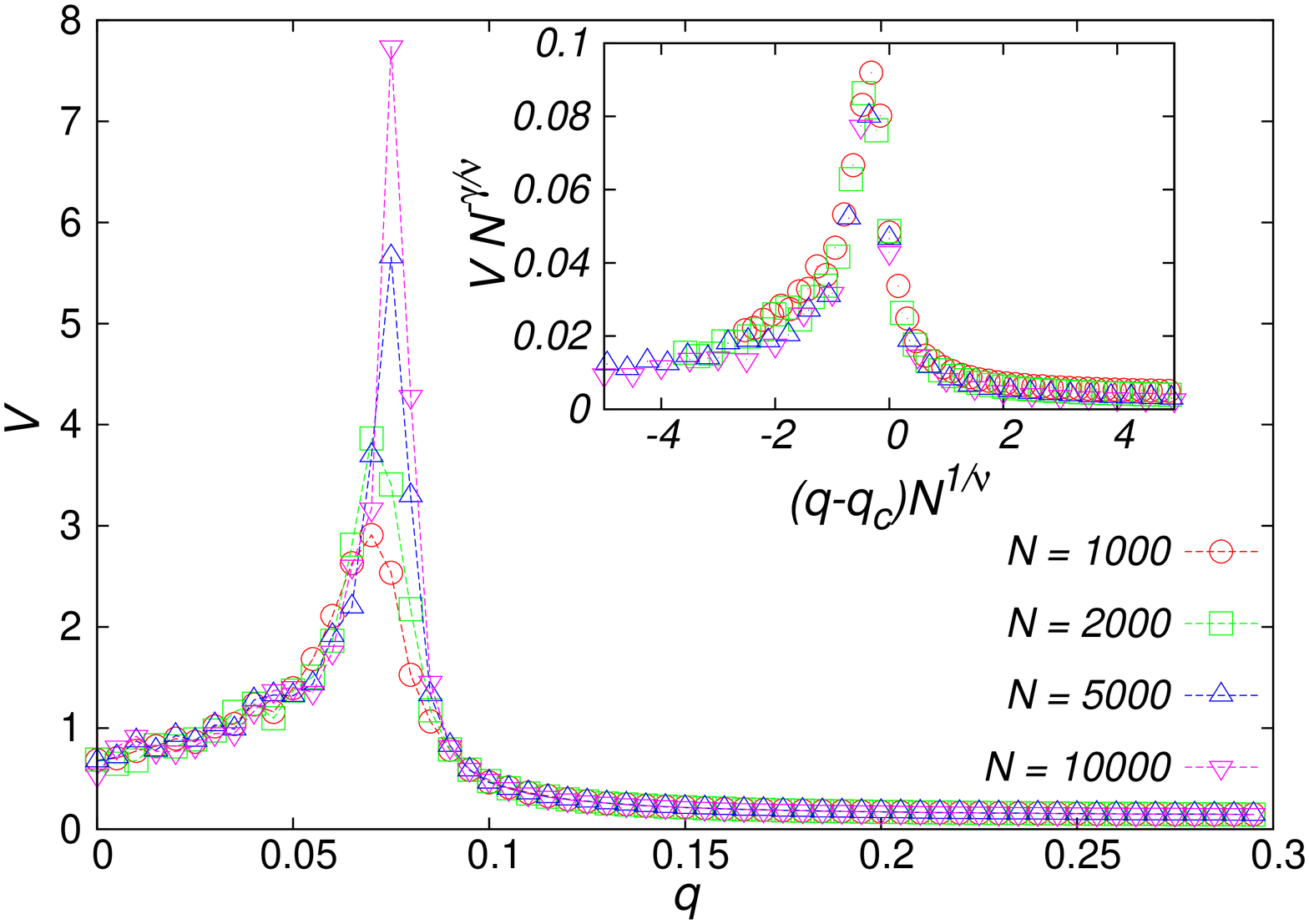}
\end{center}
\caption{Finite-size scaling analysis of the transition for $w=0.2$ and $p=0.0$. 
The best data collapse  was found for $q_{c}\approx 0.079$, $\beta\approx 1/2$, $\gamma\approx 1$ and $\nu\approx 2$.}
\label{fig6}
\end{figure}

In addition, one observes that the critical values $q_{c}(w)$ 
are much smaller than in Sec.~\ref{case1},  
where   $c_{i}=1$ for all $i$ but negative pairwise interactions $\mu_{ij}$ (i.e., $p>0$) were allowed. 
We estimated the values of the critical points $q_{c}(w)$ 
by means of the crossing of the Binder cumulant curves, 
as well as the critical exponents, as illustrated in Figure~\ref{fig6} for the 
case $w=0.2$. 
These exponents are the same as the ones of the previous section, 
namely $\beta\approx 1/2$, $\gamma\approx 1$ and $\nu\approx 2$. 
Thus, the universality class of the model is not changed by the presence of disorder 
in the convictions, as expected.

%%%%%%%%%%%%%%%%%%%%%%%%%%%%%%%%%%%%%%%%%%%%%%%%%%%%%%%%%%%%%%%%%%%%%%%%%%
%%%%%%%%%%%%%%%%%%%%%%%%%%%%%%%%%%%%%%%%%%%%%%%%%%%%%%%%%%%%%%%%%%%%%%%%%%%

\begin{figure}[h]
\begin{center}
\vspace{0.5cm}
\includegraphics[width=0.3\textwidth,angle=270]{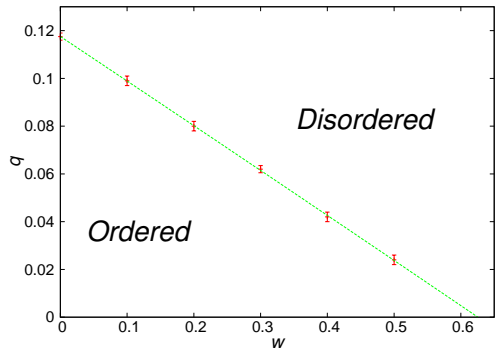}
\end{center}
\caption{Phase diagram of the model  in the plane $q$ (probability of independence) versus $w$ (probability of negative convictions), when $p=0$. The symbols are the numerical estimates of the transition points, whereas the dashed line is obtained by substituting $p$ by $w$ in Eq.~(\ref{eq5}), as explained in the text. The error bars were estimated from the FSS analysis.}
\label{fig7}
\end{figure}

Figure~\ref{fig7} shows the resulting phase diagram in the plane $q$ versus $w$, 
using the values of $q_{c}(w)$ obtained from FSS.
The boundary separates the ordered and disordered phases. 
Following the same analysis of Sec.~\ref{case1}, we estimated the 
qualitative behavior of the order-disorder frontier by means of Eq.~(\ref{eq5}), 
with the change $p\to w$.  
The result is plotted in Figure~\ref{fig7} together with the critical points obtained 
from FSS analysis. 
One can see a good agreement between the data and the qualitative frontier. 
In this case,  $c\approx 0$ and the extreme critical values   are 
$q_{c}(w=0)=b\approx 0.12$ and $w_{c}(q=0)=-b/a\approx 0.63$.

%%%%%%%%%%%%%%%%%%%%%%%%%%%%%%%%%%%%%%%%%%%%%%%%%%%%%%%%%%%%%%%%%%%%%%%%%%
%%%%%%%%%%%%%%%%%%%%%%%%%%%%%%%%%%%%%%%%%%%%%%%%%%%%%%%%%%%%%%%%%%%%%%%%%%%

\begin{figure}[h]
\begin{center}
\vspace{0.5cm}
\includegraphics[width=0.3\textwidth,angle=270]{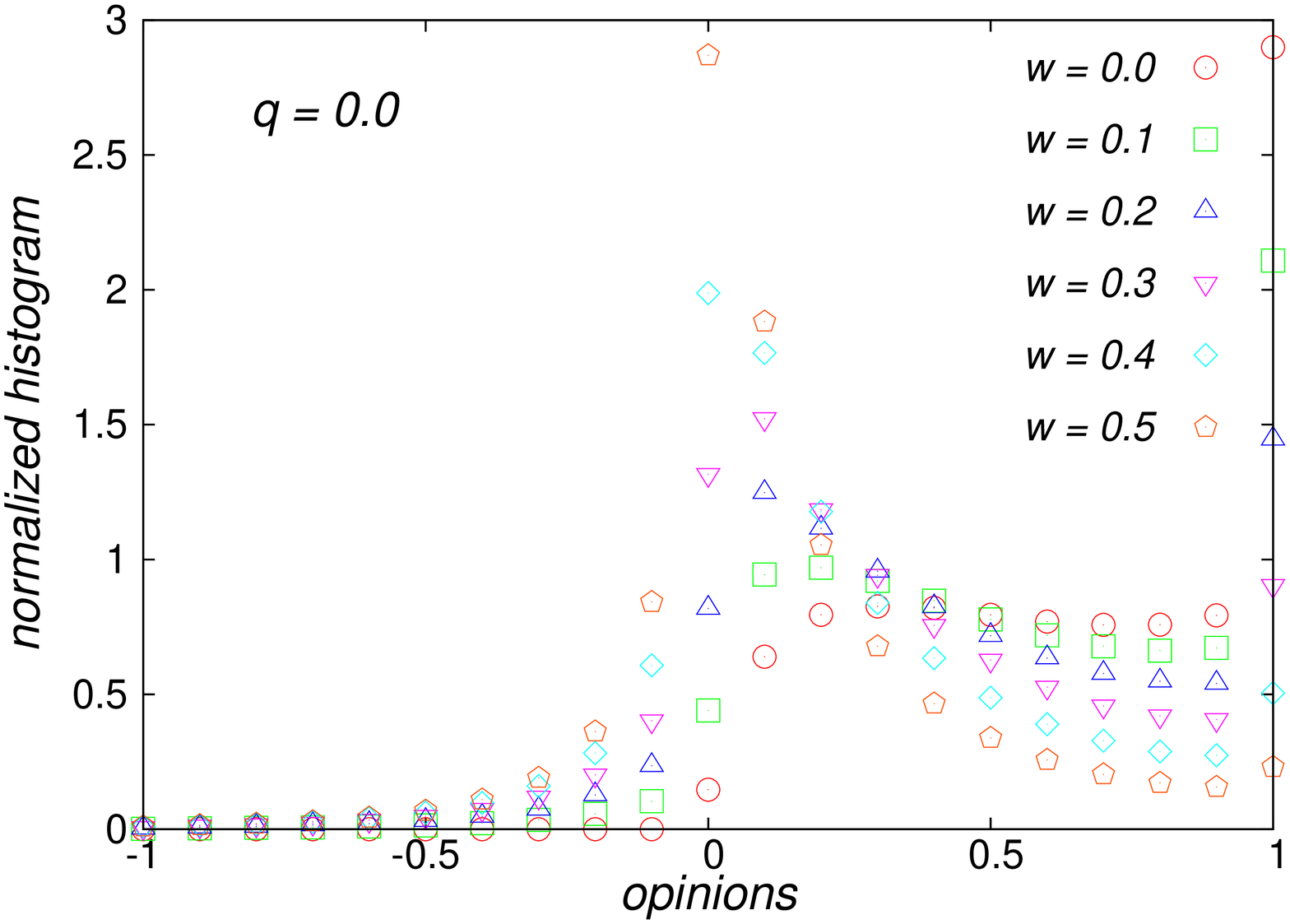}
\hspace{0.5cm}
\includegraphics[width=0.3\textwidth,angle=270]{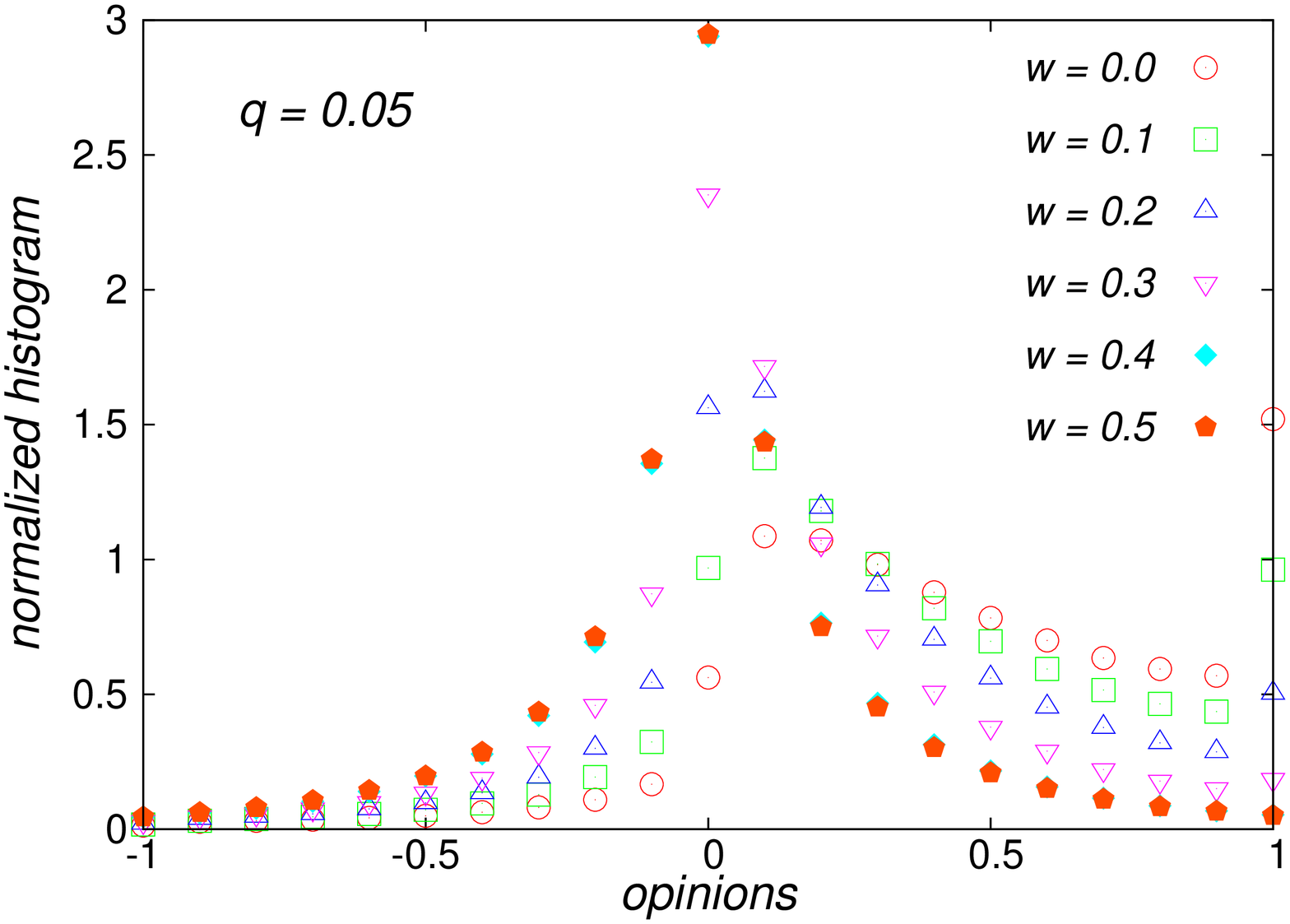}
\\
\vspace{1.0cm}
\includegraphics[width=0.3\textwidth,angle=270]{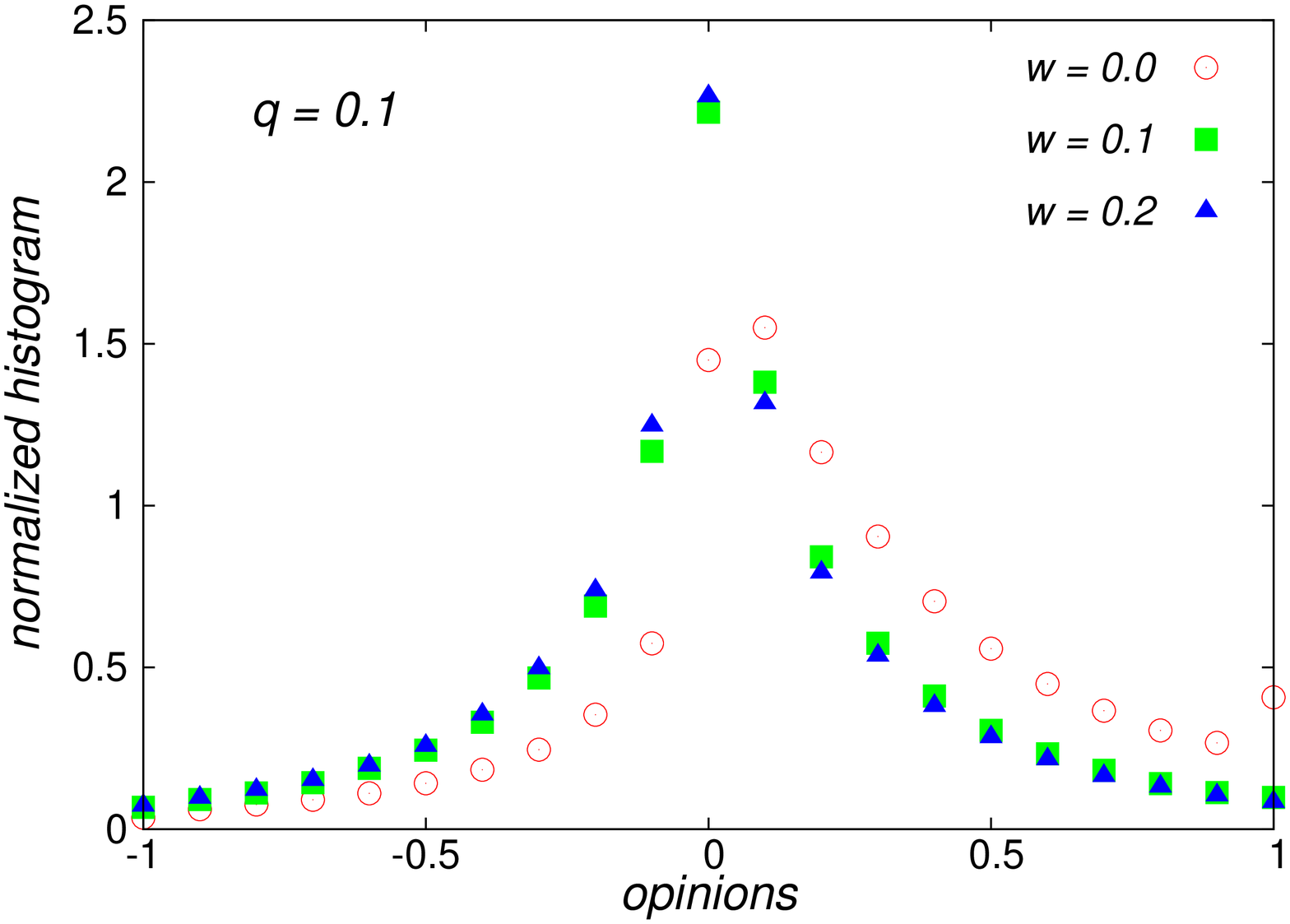}
\end{center}
\caption{Normalized histograms of opinions, taken at the steady state, 
for $q=0.0$ (top, left pannel), $q=0.05$ (top, right pannel) and $q=0.1$ (bottom), 
for several values of $w$ (and $p=0$). 
Empty (full) symbols represent ordered (disordered) states. 
The population size is $N=10^{4}$, and data are accumulated over $100$ independent simulations, as explained in the text.}
\label{fig8}
\end{figure}

 To understand the nature of the disordered phase, also in this case 
we obtained the distribution of opinions in the steady state as done in Sec.~\ref{case1}. 
In Figure~\ref{fig8} we show the normalized histograms for several values of $q$ and $w$. 
Each histogram  is built 
from $100$ independent simulations, for population size $N=10^{4}$, as in Sec.~\ref{case1}. 
%also the result of a single realization of the dynamics. 
Notice that the distributions are 
completely different from those in Figure~\ref{fig4}, both when the system is 
in the ordered and disordered phases. 
For values of the parameters in the disordered phase, i.e. at the right of critical line 
in Figure~\ref{fig7}, the plot is not flat in the center, as it was in the case of Figure~\ref{fig4}, 
but the distribution becomes symmetrically concentrated around the neutral state $o=0$ where there is a peak. 
That is,  the population becomes essentially neutral and opposite opinions are balanced.
When we move away to the left of the critical line in Figure~\ref{fig8}, 
the distribution of opinions loses symmetry, in such a way 
that one of the sides of the debate
(the negative opinions in the example of the figure)  tends to disappear,  
while the opposite (positive) side tends to become uniform, 
except for the condensation peak at $o=1$, that is the distribution becomes bimodal. 
This effect is more pronounced in the absence of negative convictions ($w=0.0$), 
which represents the farther points from the frontier, at fixed $q$.

\begin{figure}[t]
\begin{center}
\vspace{0.5cm}
\includegraphics[width=0.3\textwidth,angle=270]{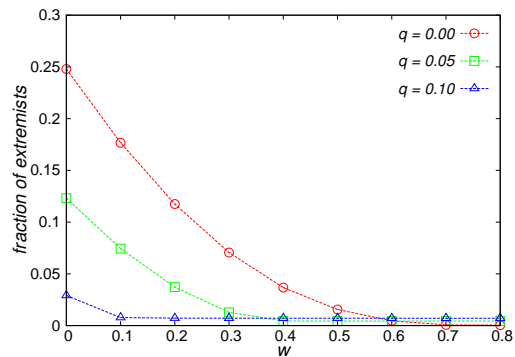}
\end{center}
\caption{Average fraction of agents with extreme opinions $|o|=1$ as a function of $w$ (probability of negative convictions) for typical values of $q$ (probability of independence), when $p=0$. One can see that the number of extremists in the population can be reduced (even to zero) for sufficiently large values of $w$. Population size is $N=10^{4}$ and averages were performed over $100$ simulations.}
\label{fig9}
\end{figure}

For increasing values of $w$, the fraction of extremists decreases 
(this effect was quantified in Figure~\ref{fig9}), 
as well as the fraction of agents with moderate opinions, while 
negative opinions emerge in the population.

From other viewpoint, the absence of negative convictions ($w=0$) 
leads to the emergence of extremists in the population, 
and the introduction of volatile agents with negative convictions 
makes moderate and indifferent opinions dominant, 
which seem realistic features of the model. 
In addition, if  the independent behavior becomes more frequent (increasing $q$), 
 the PDFs become more symmetric and there are less agents sharing extreme opinions. 
See Figure~\ref{fig9}, where  we exhibit the fraction of extremists ($|o|=1$) 
as a function of $w$ for  three different values of $q$. 
Notice that the increase of independent attitudes (increasing $q$) 
tends to reduce  the fraction of extremists in the population.
The increase of volatile attitudes (increasing $w$) favors the emergence of moderate opinions 
and  also reduces condensation at the extreme positions. 
Indeed, the increase of volatile individuals  reduces, even extinguishes, 
the fraction of extremists and 
also prevents the arrival to consensus.

Finally, let us comment that when both $\mu_{ij}$ and $c_i$ are allowed to take negative values 
i.e., $p,w>0$, 
the phenomenology is similar to that discussed in this section, although the domain of the ordered 
phase shrinks (not shown, as far as there are not new qualitative features).

Therefore, the opinion patterns observed in this section can be attributed to the joint 
heterogeneity  in the convictions and  in the interactions, be them all positive or not. 
When individuals have all the same conviction $c_i=1$, or the same interaction strength $\mu_{ij}=1$, 
then, patterns of the type observed in Figure~\ref{fig4} emerge. 
The substitution of  $c_i=1$ by positive heterogeneous convictions, like in the case $w=0$, 
is enough to yield  patterns of opinions similar to those shown in  Figure~\ref{fig8}, when we vary $p$ instead of $w$.
%

%%%%%%%%%%%%%%%%%%%%%%%%%%%%%%%%%%%%%%%%%%%%%%%%%%%%%%%%%%%%%%%%%%%%%%%%%%
%%%%%%%%%%%%%%%%%%%%%%%%%%%%%%%%%%%%%%%%%%%%%%%%%%%%%%%%%%%%%%%%%%%%%%%%%%%

%%%%%%%%%%%%%%%%%%%%%%%%%%%%%%%%%%%%%%%%%%%%%%%%%%%%%%%%%%%%%%%%%%%%%%%%%%
%%%%%%%%%%%%%%%%%%%%%%%%%%%%%%%%%%%%%%%%%%%%%%%%%%%%%%%%%%%%%%%%%%%%%%%%%%%

% ############################################################################

\section{Final Remarks}

\qquad In this work, we have studied a kinetic model of opinion formation with continuous states. 
We considered pairwise interactions among randomly chosen agents, together with the possibility of independent behaviors that 
occur  with probability $q$.  
Two sources of heterogeneity were included:  random competitive couplings (interaction term)   
and random convictions (self-interaction term), that can take negative values with probabilities $p$ and $w$, 
respectively.   We characterized the critical transitions, 
as well as the emerging phases, in terms of moderate and extremist individuals.

Parameters  $p$, $w$ and $q$ rule microscopic disorder in the system, and their increase triggers  
transitions from collective order to disorder.
These transitions are characterized by mean-field exponents and their critical curves are shown in Figures ~\ref{fig3} and ~\ref{fig7}. 

When only one source of heterogeneity is present, like in the case of Sec.~\ref{case1}, 
where the convictions are given by $c_i=1$ for all $i$, but there is a probability $p>0$ of negative couplings, 
patterns of opinions like those shown in Figure~\ref{fig4} occur. 
Qualitatively similar patterns arise when the pairwise interactions are $\mu_{ij}=1$ for all $i,j$ but convictions are random.
In the disordered phase, opinions are evenly distributed. 
In the ordered ones,   extremists dominate.
The steady states show the presence of a large number of extremists with opinions either $o=1$ or $o=-1$. 
The fraction of these extremists decreases for increasing values 
of the independence parameter $q$, and the system tends to become disordered, indicating 
that independent attitudes reduce extreme opinions.

Different patterns emerge when there are two sources of heterogeneity, with negative values or not, 
like in the case of  Sec.~\ref{case2}, 
where the couplings and the convictions are random. 
In the disordered phase, opinions are not evenly distributed, but moderate opinions dominate, with a peak around the neutral state. In the ordered phases, one of the opinion sides disappears. When a second source of heterogeneity is present, the increase of volatile individuals (increasing $w$) or of dissent interactions (positive $p$), reduces, even extinguishes, the fraction of extremists and also prevents the arrival to consensus. 

The models considered in this work contribute to the analysis of the role of the diverse features that characterize individuals and their interactions. 
As can be seen in this paper, each modification yields phase diagrams of order parameters that seem similar, however a quite different phenomenology arises in terms of the distribution of opinions. This hinders unification, justifying a separate analysis of different versions of the model. These versions  shed light on the origin and role of undecided individuals and extremism uprise, and the effect of realistic characteristics like conviction and independence in the dynamics of opinion formation and evolution. In particular, the models contribute to understand the circumstances which favor the emergence and development of extremism in a fraction of the population, relating that emergence with the existence of individuals with strong positive convictions in the population. On the other hand, the presence of negative convictions, that represent volatile individuals that have a propensity to change mind, as well as nonconformity represented in the model by the independent behavior, lead to the dominance of moderate individuals.

% ############################################################################

\section*{Acknowledgments}

The authors acknowledge financial support from the Brazilian funding agencies CNPq and FAPERJ.


\section*{References}
\begin{thebibliography}{30}

 
\bibitem{loreto_rmp}
C. Castellano, S. Fortunato, V. Loreto, Rev. Mod. Phys. 81, 591 (2009).

\bibitem{galam_book}
S. Galam, \textit{Sociophysics: A Physicist's Modeling of Psycho-political Phenomena} (Springer, Berlin, 2012).

\bibitem{sen_book}
P. Sen, B. K. Chakrabarti, \textit{Sociophysics: an introduction} (Oxford University Press, Oxford, 2013).

\bibitem{galam_review}
S. Galam, Int. J. Mod. Phys. C 19, 409 (2008).

\bibitem{pawel}
P. Sobkowicz, J. Artif. Soc. Soc. Simul. 12(1):11 (2009).

\bibitem{voter} T. M. Liggett, \textit{Stochastic Interacting Systems: Contact, 
Voter and Exclusion Processes} (Springer-Verlag, 1999).

\bibitem{sznajd} K. Sznajd-Weron, J. Sznajd, Int. J. Mod. Phys. C 11, 1157 (2000).

\bibitem{majority} P. Chen, S. Redner, J. Phys. A Math. Gen. 7239, 8 (2005).

\bibitem{sen}
P. Sen, Phys. Rev. E 83, 016108 (2011).

\bibitem{lccc}
M. Lallouache, A. S. Chakrabarti, A. Chakraborti, B. K. Chakrabarti, Phys. Rev. E 82, 056112 (2010).

\bibitem{biswas}
S. Biswas, A. Chatterjee, P. Sen, Physica A 391, 3257 (2012).


\bibitem{galam}
S. Galam, Physica A 333 (2004) 453-460.

\bibitem{lalama}
M. S. de la Lama, J. M. Lopez, H. S. Wio, Europhys. Lett. 72 (2005) 851-857.

\bibitem{sznajd_indep1}
K. Sznajd-Weron, M. Tabiszewski, A. M. Timpanaro, Europhys. Lett. 96 (2011) 48002;

\bibitem{sznajd_indep2}
P. Nyczka, K. Sznajd-Weron, J. Cislo, Phys. Rev. E 86 (2012) 011105.

\bibitem{sznajd_indep3}
P. Nyczka, K. Sznajd-Weron, J. Stat. Phys. 151 (2013) 174-202.

\bibitem{nuno_indep}
N. Crokidakis, Phys. Lett. A 378, 1683 (2014).

\bibitem{javarone}
M. A. Javarone, Physica A 414, 19 (2014).

\bibitem{bagnoli}
F. Bagnoli, R. Rechtman, Phys. Rev. E 92, 042913 (2015).


\bibitem{willis}
R. H. Willis, Sociometry 26, 499 (1963).

\bibitem{nail}
P. R. Nail, G. MacDonald, Psychol. Bull. 126, 454 (2000).


%%%%%%%%%%%%%%%%%%%%%%%%%%

% convictions:
 

\bibitem{deffuant2}
G. Deffuant, J. Artif. Soc. Soc. Simul. 9(3):8 (2006). %, available at http://jasss.soc.surrey.ac.uk/9/3/8.html.

\bibitem{nuno_celia}
N. Crokidakis, C. Anteneodo, Phys. Rev. E 86, 061127 (2012).

\bibitem{brugna}
C. Brugna, G. Toscani, Phys. Rev. E 92, 052818 (2015).

\bibitem{nuno_pmco_jstat}
N. Crokidakis, P. M. C. de Oliveira,  J. Stat. Mech. P11004 (2011).

\bibitem{xiong}
F. Xiong, Y. Liu, J. Zhu, Entropy 15, 5292 (2013).

\bibitem{nuno_jstat}
N. Crokidakis, J. Stat. Mech. P07008 (2013).

\bibitem{sibona}
G. R. Terranova, J. A. Revelli, G. J. Sibona, Europhys. Lett. 105, 30007 (2014).

\bibitem{marlon}
M. Ramos, J. Shao, S. D. S. Reis, C. Anteneodo, J. S. Andrade, S. Havlin, H. A. Makse, Scientific Reports 5, 10032 (2015).


 \bibitem{victor} 
N. Crokidakis, V. H. Blanco,  C. Anteneodo, Phys. Rev. E 89, 013310 (2014).



\bibitem{fan}
K. Fan, W. Pedrycz, Physica A 436, 87 (2015).

%%%%%%%%%%%%%%%%%%%%%%%%%
% models with continuous opinions

\bibitem{hk}
R. Hegselmann, U. Krause, J. Artif. Soc. Soc. Simul. 5(3), (2002) %, available at http://www.soc.surrey.ac.uk/JASSS/5/3/2.html

\bibitem{deffuant}
G. Deffuant, D. Neau, F. Amblard, G. Weisbuch, Advs. Complex Syst. 03, 87 (2000).

\bibitem{deffuant3}
G. Deffuant, F. Amblard, G. Weisbuch, T. Faure, J. Artif. Soc. Soc. Simul. 5(4):1 (2002).

\bibitem{lorenz}
J. Lorenz, Int. J. Mod. Phys. C 18, 1819 (2007).


\bibitem{coda}
A. C. R. Martins, Int. J. Mod. Phys. C 19, 617 (2008).


\bibitem{wu}
Y. Wu, Y. Hu, X. He, Int. J. Mod. Phys. C 24, 1350080 (2013).



\bibitem{binder}
K. Binder, Z. Phys. B 43, 119 (1981).





\end{thebibliography}
\end{document}